\documentclass[twoside,web]{ieeecolor}
\usepackage{generic}
\usepackage{amsmath,amssymb,amsfonts}
\usepackage{algorithmic}
\usepackage{graphicx}
\usepackage{textcomp}
\usepackage{xcolor}
\usepackage[version=4]{mhchem}
\usepackage[
    backend=biber,
    style=ieee,
    citestyle=ieee-comp,
    maxnames=99,
    minnames=99,
    doi=true,
    url=false,
    isbn=false,
]{biblatex}

\def\BibTeX{{\rm B\kern-.05em{\sc i\kern-.025em b}\kern-.08em
    T\kern-.1667em\lower.7ex\hbox{E}\kern-.125emX}}
\begin{document}
\title{Gate-defined quantum dot and charge noise analysis in a Forksheet transistor
}
\author{Dominic~Leclerc, Claude~Rohrbacher,
         Alexis~Morel, Joffrey~Rivard, 
         Romain~Ritzenthaler,
        Christian~Lupien, Hans~Mertens, Naoto~Horiguchi
        and~Eva~Dupont-Ferrier
\thanks{D. Leclerc, C. Rohrbacher, J. Rivard, C. Lupien and E. Dupont-Ferrier are with Institut quantique and D\'epartement de Physique,
Universit\'e de Sherbrooke, Sherbrooke,
Qu\'ebec, J1K 2R1, Canada. (e-mail: dominic.leclerc2@usherbrooke.ca)}
\thanks{R. Ritzenthaler, H.Mertens and N. Horiguchi are with IMEC, 3001 Leuven, Belgium.(e-mail: Romain.Ritzenthaler@imec.be)}}

\maketitle

\begin{abstract}
    We report the cryogenic characterization of a 14-nm nMOS Forksheet transistor down to 10 mK. We demonstrate that the device retains transistor operation with improved characteristics at cryogenic temperatures, while also forming a gate-defined quantum dot that can be tuned to the single-electron regime. In the quantum-dot regime, we perform low-frequency charge-noise measurements across different charge occupations and observe an average charge-noise amplitude of $\langle S_0\rangle = 10\;\mu\mathrm{eV}/\sqrt{\mathrm{Hz}}$ at 1 Hz, with a minimum measured value of $3.3\;\mu\mathrm{eV}/\sqrt{\mathrm{Hz}}$. These results highlight the potential of Forksheet technology for the monolithic cointegration of quantum and classical CMOS functionalities in future quantum processors.
\end{abstract}


\section{Introduction}
\IEEEPARstart{C}{MOS} technology has shown great promise recently for scaling up quantum computers by implementing cryogenic CMOS electronics (Cryo-CMOS) to control quantum processors \cite{gonzalez-zalba_scaling_2021,ruffino_cryo-cmos_2021,bartee_spin-qubit_2025,hrl_digitally_2026}. In parallel, CMOS compatible spin-qubit devices have demonstrated excellent performance, achieving single-qubit and two-qubit gate fidelities above 99 $\%$ \cite{xue_quantum_2022,Noiri2022,tanttu_assessment_2024,steinacker_industry-compatible_2025}, while maintaining high performance at temperatures exceeding 1 K \cite{camenzind_hole_2022, huang_high-fidelity_2024}. Together, these advances pave the way toward fully cointegrated quantum processors combining qubits and classical control electronics on the same chip. In particular, emerging gate-all-around (GAA) technology nodes such as nanosheets are especially attractive for this purpose, as they can operate both as classical transistors and as gate-defined quantum dots at cryogenic temperatures \cite{rohrbacher2023dual,rohrbacher_nanosheet_2025}. Forksheet devices are envisioned as candidate devices for post-nanosheet nodes \cite{weckx_novel_2019}. In a Forksheet device, the separation between N- and P-FETs is done with a dielectric wall, enabling further area optimization or maximization of sheet width (and consequently current) for iso-standard cell area. A previous study at 4 K performed on an array of 800 nMOS Forksheets in parallel demonstrated improved performance at cryogenic temperatures \cite{asanovski_characterization_2024}, highlighting the potential of this technology for cryo-CMOS applications. In this work, we study an individual 14-nm Forksheet nMOS  transistor fabricated at IMEC in a 300~mm integration line for cryogenic quantum computing applications. We first demonstrate transistor operation by performing transfer-characteristic measurements from 300 K down to 10 mK. We then demonstrate the formation of a gate-defined quantum dot in the Forksheet device and characterize its charge noise by using the quantum dot as a charge sensor.

\section{Device and method}
\begin{figure}[ht]
\centering
\includegraphics[width = 0.6\columnwidth]{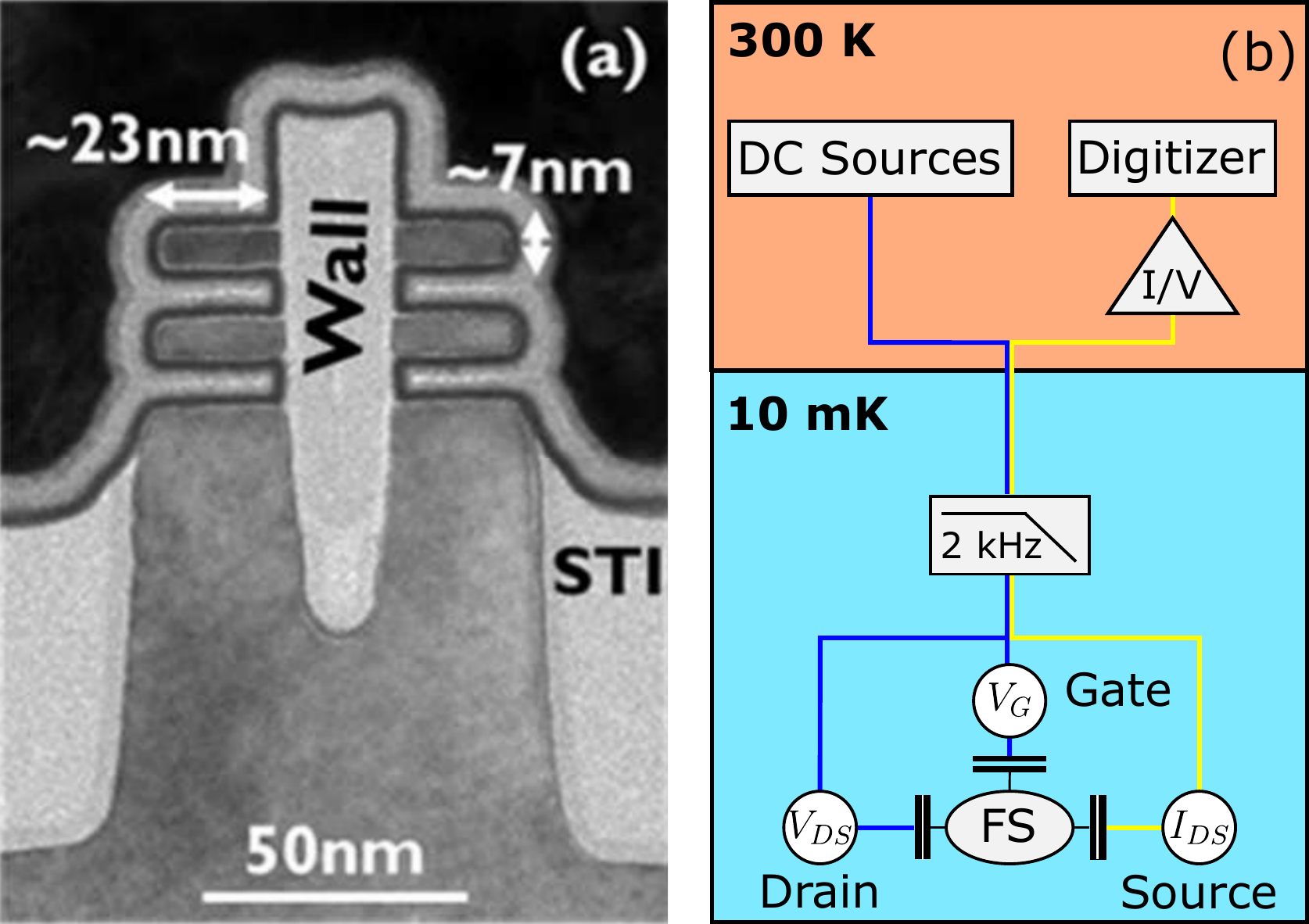}
\caption{\textbf{a)} TEM image across the conduction channel of the Forksheet device from \cite{ritzenthaler_comparison_2021}. \textbf{b)} Illustration of the measurement setup with the measured Forksheet (FS) device at 10 mK.}
\label{fig:1}
\end{figure}

\begin{figure}[ht]
\centering
\includegraphics[width = 0.9\columnwidth]{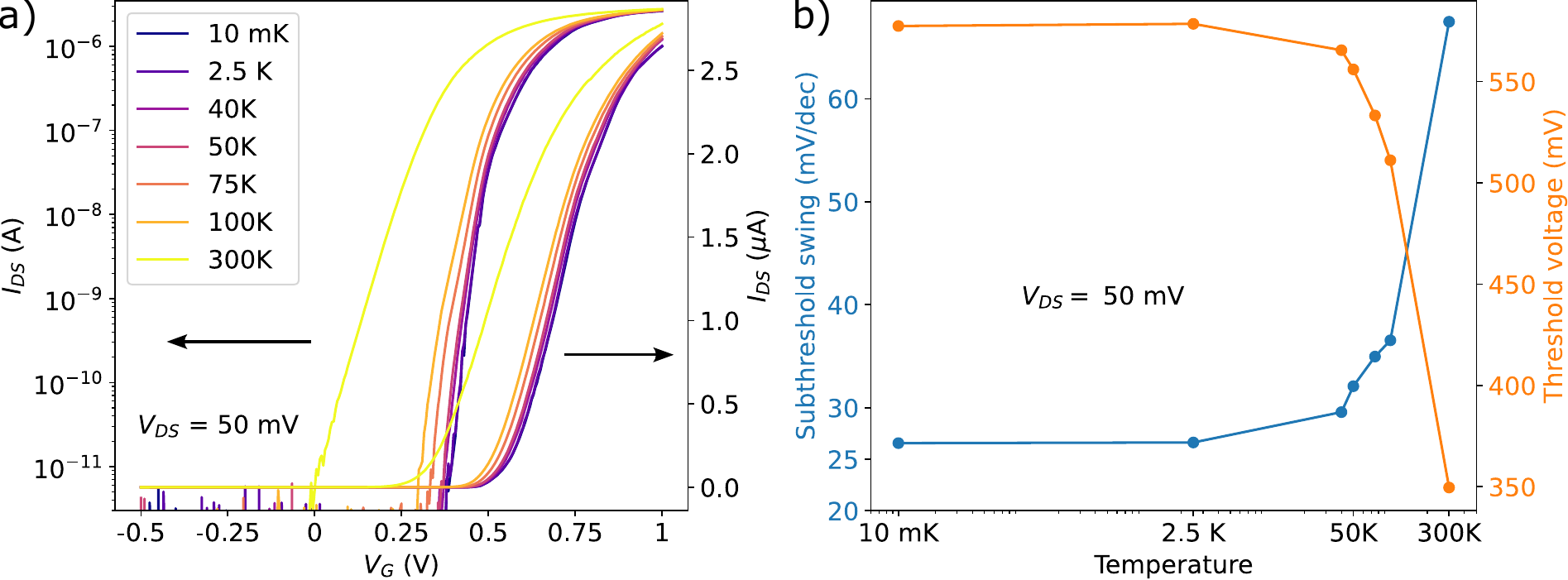}
\caption{Transfer characteristics of the Forksheet device as a function of temperature. \textbf{a)} Drain current $I_{DS}$ as a function of gate voltage $V_G$ at different temperatures, shown on a logarithmic scale (left axis) and a linear scale (right axis). \textbf{b)} Extracted subthreshold swing and threshold voltage as a function of temperature.}
\label{fig:2}
\end{figure}

The measured device consists of two lateral Nanosheet devices separated by an inner \ce{SiN} wall, all connected by the same gate of layout length $L_G=$ 14 nm as shown in Fig. \ref{fig:1} \textbf{a)}. Further details on the device fabrication process can be found in \cite{ritzenthaler_comparison_2021,mertens2021forksheet}. The sample is fixed on the mixing plate of a Bluefors LD250 dilution refrigerator with a base temperature of 10 mK. The sample is wire-bonded to a PCB and connected to a homemade 2nd order low-pass RC filter with a cutoff frequency of 2 kHz, also fixed on the mixing plate to reduce noise and lower electronic temperature, as illustrated in Fig. \ref{fig:1} \textbf{b)}. DC biases are applied using an iTest Bilt BE2142 low-noise voltage source. The current is measured using a Basel SP 983c transimpedance amplifier at room temperature, followed by voltage measurement with a Keysight 34465A digital multimeter. For noise measurements, a Data Translation DT9837C data acquisition card is used with a sampling frequency of 2 kHz.
Transistor $I_{DS}-V_G$ curves from room temperature down to 10 mK at $V_{DS}=50$ mV are shown in Fig. \ref{fig:2} \textbf{a)}. The threshold voltage $V_{th}$ is extracted using the linear regression around the maximum of transconductance \cite{Ortiz-Conde2002} and shown in Fig. \ref{fig:2} \textbf{b)}. The evolution of $V_{th}$ with temperature reveals an increase of 228 mV from 300 K to 10 mK. This increase is primarily attributed to bandgap widening and scaling of the Fermi–Dirac distribution at cryogenic temperatures \cite{Bohuslavskyi2019_bandtail}. We also extracted the subthreshold swing $SSw$ at $I_{DS}=10$ pA in Fig. \ref{fig:2} \textbf{b)} showing a decrease of $SSw$ with temperature followed by a saturation at 50 K around 26 mV/dec. Such saturation, commonly observed across various CMOS technologies \cite{beckers_characterization_2019,Han2021,Incandela2017,Sekiguchi2021, rohrbacher_nanosheet_2025}, is often associated with an exponential band-tail in the conduction band \cite{Bohuslavskyi2019_bandtail}. In the present device, however, the saturation is more likely caused by direct source–drain tunnelling owing to the very short channel of this Forksheet device \cite{kao_subthreshold_2020, Sekiguchi2021}.

\section{Quantum dot operation}
We then characterize the same device in the quantum-dot regime by operating it at low temperature, below the threshold voltage and at low source-drain bias $V_{DS}$. To observe quantum effects such as Coulomb blockade, two conditions must be satisfied. First, the tunnel barriers between the conduction channel and the source and drain reservoirs must be sufficiently opaque to ensure well-defined charge quantization. Second, the charging energy, defined as $E_C=e^2/C_\Sigma$, where $C_\Sigma$ is the total capacitance of the device (sum of source, drain and gate capacitances), must exceed the thermal energy $k_BT$. In the present Forksheet architecture, the tunnel barriers arise from the separation between the highly doped source and drain regions and the conduction channel, as the dopants do not extend into the channel region \cite{ritzenthaler_comparison_2021}.
\begin{figure}[ht]
\centering
\includegraphics[width =  0.9\columnwidth]{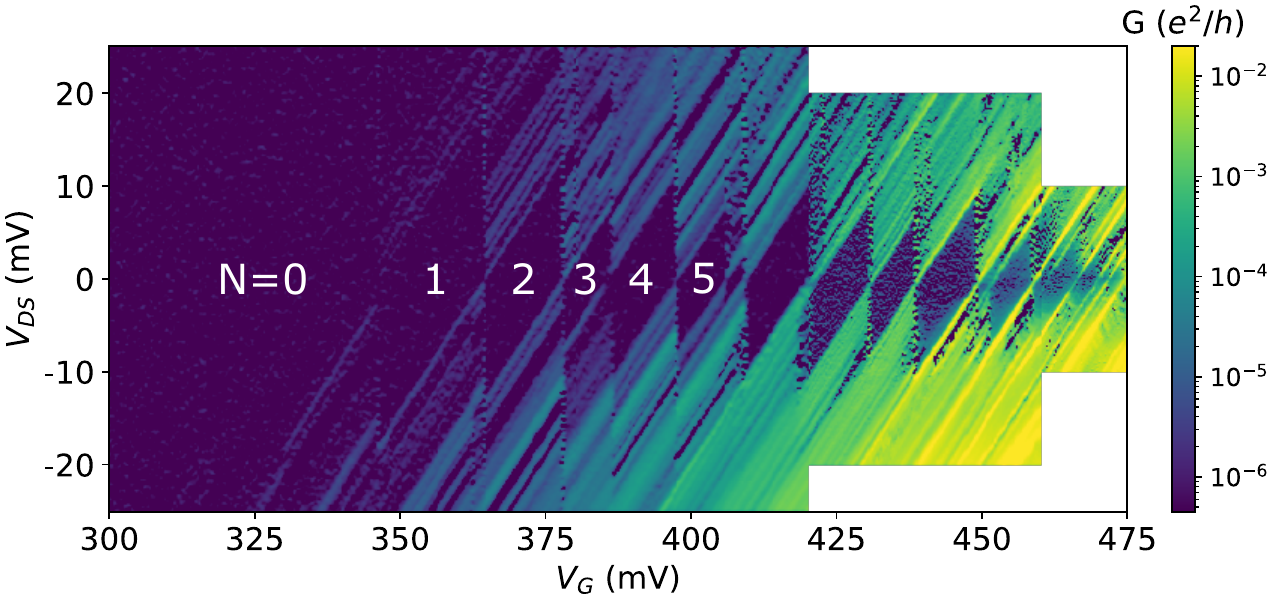}
\caption{Measured stability diagram of the Forksheet device at $T=10$ mK, with the electron filling identified from $N=0$ to $N=5$. The color scale represents the differential conductance $G=\partial I_{DS}/\partial V_{DS}$ in units of $e^2/h$, computed numerically.}
\label{fig:3}
\end{figure}
A stability diagram measured at a base temperature of 10 mK is shown in Fig. \ref{fig:3}, exhibiting the characteristic Coulomb diamonds associated with a gate-defined quantum dot formed in one of the Forksheet channel. From the height and width of the Coulomb diamonds, we extract the charging energy $E_C$ and the lever arm, defined as $\alpha=eC_G/C_\Sigma$, where $C_G$ is the gate capacitance. Over the range of $V_G$ between 345 mV and 450 mV, including the first electron filling of the quantum dot, we obtain $\langle E_C\rangle=9.7\pm3.8$ meV and $\langle\alpha\rangle=0.94\pm0.04$ eV/V. This large lever arm is consistent with the tri-gate architecture of the Forksheet channel and is a key parameter for implementing fast control and readout techniques such as gate-based reflectometry \cite{frake_radio-frequency_2015,rivard_multimode_2026}. For comparison, typical lever-arm values for silicon spin-qubit quantum dots are below 0.1 eV/V \cite{freeman_comparison_2016,elsayed_low_2024} in planar architectures, while values above 0.6 eV/V have been reported for gate-all-around architectures \cite{rohrbacher2023dual,rohrbacher_nanosheet_2025}.

\section{Charge noise characterization}
\begin{figure}[h]
\centering
\includegraphics[width= 0.9\columnwidth]{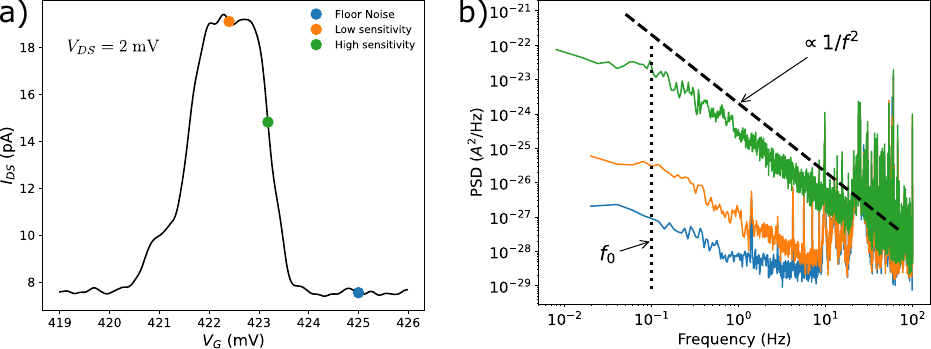}
\caption{Example of a charge-noise measurement and comparison with other noise sources. \textbf{a)} Coulomb peak at $V_{DS}=2$ mV showing the three points of comparison. \textbf{b)} Power spectral density (PSD) measured at the three points of comparison, illustrating the range of validity of the charge-noise measurement and the Lorentzian spectral shape with a characteristic frequency $f_0\sim0.1$ Hz.}
\label{fig:4}
\end{figure}

\begin{figure}[ht]
\centering
\includegraphics[width=0.9\columnwidth]{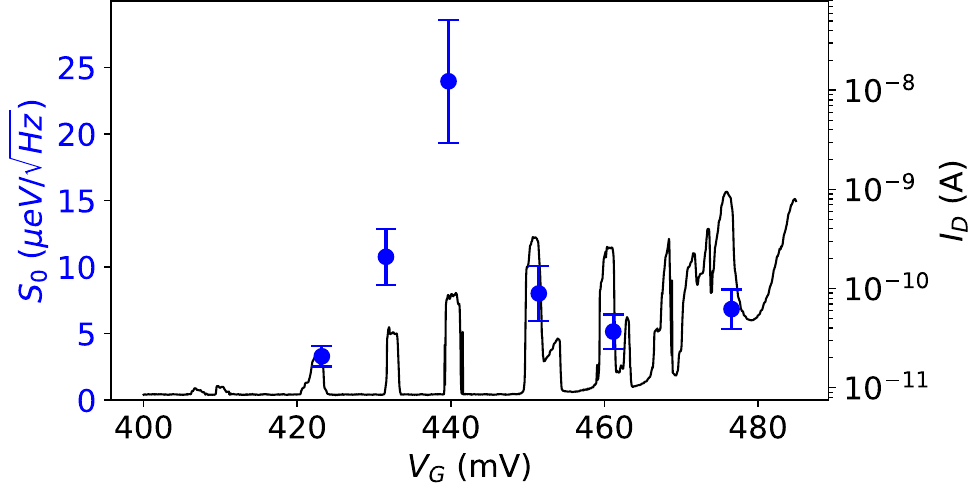}
\caption{Extracted value of charge noise $S_0$ at $f=1 Hz$ (Blue dot), over different Coulomb peaks (Black line) at $V_{DS}= 2$ mV.}
\label{fig:5}
\end{figure}

Evaluation of device charge noise is a key element for implementation of spin qubit devices since it is one of the principal limitations for the fidelity of readout and one- and two-qubit gates \cite{Yoneda2017, Kranz2020,tanttu_assessment_2024}. Charge noise in gate-defined quantum dots is primarily attributed to two-level fluctuators (TLFs) that switch between two metastable states with a characteristic frequency $f_0$ \cite{Bermeister_TLF}. Each TLF exhibits a Lorentzian frequency response, while a broad distribution of TLFs gives rise to the well-known $1/f^\gamma$ noise spectrum at low frequencies, with $\gamma$ typically ranging from 0.5 to 2 \cite{Kranz2020}. These TLFs are believed to arise from defects near the interface of different materials, notably the \ce{Si}/\ce{SiO2} interface and the \ce{SiO2}/high-$k$ dielectric interface in CMOS devices. Since these TLFs are typically located near the interface between the conduction channel and the gate dielectric, their fluctuations can manifest as voltage noise or variation in the electrostatic potential of the quantum dot. To measure charge noise in our device, we use the gate-defined quantum dot as a sensitive charge sensor to detect changes in its electrostatic environment \cite{connors_low-frequency_2019,Kranz2020}. The charge noise is characterized by measuring the current power spectral density (PSD), expressed in $A^2/\mathrm{Hz}$, at the point of maximum sensitivity to the gate voltage along a Coulomb peak, as shown in Fig. \ref{fig:4}.
To ensure that the measured noise is intrinsic to the quantum dot and not dominated by setup noise or other sources, such as thermal noise and tunnel-barrier fluctuations, spectra are also acquired at the top of the Coulomb peak, where the charge sensitivity is low, and in the Coulomb blockade regime (Fig. \ref{fig:4}).  This measurement is repeated over different Coulomb peaks for statistics and to probe potential charge occupancy dependence of the charge noise. All the noise spectra measured at the sensitive point of a Coulomb peak show a strong Lorentzian shape (Fig. \ref{fig:4} \textbf{b)} in green) rather than the usual $1/f$. This can be attributed to the signature of a single TLF dominating the noise in this frequency range. The identified TLF has a characteristic frequency of approximately 0.1 Hz. From these measurements, the current noise $S_I(f)$ at the point of maximum Coulomb-peak sensitivity is converted into chemical-potential noise in $\mathrm{eV^2/Hz}$ using:
\begin{equation}
    |S_\mu (f)|^2 = \frac{\alpha^2|S_I (f)|^2}{\left(\partial I/\partial V_{G}\right)^2},
\end{equation}
where $\alpha$ is the lever arm and $\partial I/\partial V_{G}$ is extracted at the maximum slope of the Coulomb peak. To compare the noise level with values reported in the literature, we extract the noise amplitude at 1 Hz, $S_0=S_\mu(f=1\;\text{Hz})$, expressed in $\mu\text{eV}/\sqrt{\text{Hz}}$, by fitting the noise spectrum with a combination of a $1/f^\gamma$ component and a Lorentzian function. The extracted $S_0$ values across different Coulomb peaks are presented in Fig. \ref{fig:5}, yielding an average value of $\langle S_0\rangle = 10\;\mu\text{eV}/\sqrt{\text{Hz}}$ with a minimum of 3.3 $\mu\text{eV}/\sqrt{\text{Hz}}$ and no clear dependence on charge occupancy. This average value of $S_0$ lies at the upper end of the range reported for state-of-the-art silicon spin-qubit devices, typically spanning $0.5$–$10\;\mu\text{eV}/\sqrt{\text{Hz}}$ \cite{elsayed_low_2024,freeman_comparison_2016,Zwerver2021, Kranz2020}. However, the observed noise level is particularly notable given the device architecture and materials. Unlike many silicon spin-qubit devices, the present device employs a high-$k$ dielectric, introducing an additional interface in close proximity to the quantum dot. High-$k$ oxides such as \ce{HfO2} can also exhibit increased defect densities due to lattice mismatch with the \ce{SiO2} interfacial layer \cite{haartman_low-frequency_2007}. Furthermore, the gate-all-around architecture places the quantum dot in proximity to the Si/oxide interface over a large portion of its perimeter, potentially increasing its exposure to nearby TLFs. Despite these potentially unfavorable conditions, the measured $\langle S_0\rangle$ remains within the range reported for state-of-the-art silicon spin-qubit devices. Importantly, this noise level is obtained using a standard CMOS gate stack and fabrication flow, without process optimization specifically targeting quantum-device performance. The Forksheet device therefore demonstrates that low charge-noise levels can be achieved in a scalable 300~mm CMOS platform with a large lever arm, highlighting its potential for scalable quantum–classical integration.

\section{Conclusion}
We demonstrated that a 14-nm nMOS Forksheet transistor exhibits improved performance at 10 mK in the classical transistor regime while also hosting a gate-defined quantum dot with a large lever arm. The charge occupation of the quantum dot can be precisely controlled and tuned down to the single-electron regime, a prerequisite for spin qubit operation. The large lever arm of the device provides efficient gate control, making it well suited for gate-based reflectometry and potentially enabling qubit readout without the necessity of an additional nearby charge sensor. Additionally, charge noise measurements in the quantum dot regime show an average noise level at 1 Hz comparable to those reported for CMOS spin qubit devices. These results demonstrate the potential of Forksheet transistors as a platform for scalable quantum–classical cointegration, combining spin qubits and cryogenic control electronics within a scalable 300~mm CMOS manufacturing process.

\printbibliography

\end{document}